\documentclass[fleqn,12pt,twoside]{article}

\usepackage{espcrc1}
\usepackage{graphicx}
\usepackage{picins}

\title{\vspace{-10mm}First Results on In-Beam $\gamma$ Spectroscopy
       of Neutron-Rich Na and Mg Isotopes at REX-ISOLDE}
\newcommand{\hsif}{\vspace{-4mm}\small}
\author{%
\small
H. Scheit\address[MPI]%
{\hsif Max-Planck-Institut f\"ur Kernphysik, Heidelberg, Germany},
O.~Niedermaier\addressmark[MPI],
M.~Pantea\address[TUD]%
{\hsif Institut f\"ur Kernphysik, Technische Universit\"at Darmstadt, Darmstadt, Germany},
F.~Aksouh\address[KUL]%
{\hsif Instituut voor Kern- en Stralingsfysica, University of Leuven, Leuven, Belgium},
C.~Alvarez\address[LMU]%
{\hsif Ludwig-Maximilians-Universit\"at M\"unchen, M\"unchen, Germany},
F.~Ames\addressmark[LMU],
T.~Behrens\address[TUM]%
{\hsif Technische Universit\"at M\"unchen, M\"unchen, Germany},
V.~Bildstein\addressmark[MPI],
H.~Boie\addressmark[MPI],
P.~Butler\address[CERN]%
{\hsif CERN, Geneva, Switzerland},
J.~Cederk\"all\addressmark[CERN],
T.~Davinson\address%
{\hsif University of Edinburgh, Edinburgh, UK}
P.~Delahaye\addressmark[CERN],
P.~Van Duppen\addressmark[KUL],
J.~Eberth\address[IKP]%
{\hsif Institut f\"ur Kernphysik, Universit\"at K\"oln, K\"oln, Germany},
S.~Emhofer\addressmark[LMU],
J.~Fitting\addressmark[MPI],
S.~Franchoo\address%
{\hsif Universit\"at Mainz, Mainz, Germany},
R.~Gernh\"auser\addressmark[TUM],
G.~Gersch\addressmark[IKP],
D.~Habs\addressmark[LMU],
R.~v.~Hahn\addressmark[MPI],
H.~Hess\addressmark[IKP],
A.~Hurst\address[OLL]{\hsif Oliver Lodge Laboratory, University of Liverpool, UK},
M.~Huyse\addressmark[KUL],
O.~Ivanov\addressmark[KUL],
J.~Iwanicki\addressmark[OLL]\thanks{current address: Heavy Ion Laboratory, Warsaw University, Warsaw, Poland},
O.~Kester\addressmark[LMU],
F.~K\"ock\addressmark[MPI],
T.~Kr\"oll\addressmark[TUM],
R.~Kr\"ucken\addressmark[TUM], 
M.~Lauer\addressmark[MPI],
R.~Lutter\addressmark[LMU],
P.~Mayet\addressmark[KUL],
M.~M\"unch\addressmark[TUM],
U.K.~Pal\addressmark[MPI],
M.~Pasini\addressmark[LMU],
P.~Reiter\addressmark[IKP],
A.~Richter\addressmark[TUD],
A.~Scherillo\addressmark[IKP]\address{\small Institut Laue-Langevin, Grenoble, France},
G.~Schrieder\addressmark[TUD],
D.~Schwalm\addressmark[MPI],
T.~Sieber\addressmark[CERN],
H.~Simon\addressmark[TUD],
O.~Thelen\addressmark[IKP],
P.~Thirolf\addressmark[LMU],
J.~van de Walle\addressmark[KUL],
N.~Warr\addressmark[IKP],
D.~Wei\ss{}haar\addressmark[IKP],
for the REX and MINIBALL collaborations}

\begin{document}
\sloppy
\maketitle

\newcommand{\hssp}{-2mm}
\begin{abstract}
\small
\vspace{\hssp}
After the successful commissioning of the radioactive beam experiment 
at ISOLDE (REX-ISOLDE) --- an accelerator for exotic nuclei produced by ISOLDE ---
first physics experiments using these beams were performed. 
Initial experiments focused on the region of deformation in the vicinity of the neutron-rich
Na and Mg isotopes.
Preliminary results show the high potential and physics opportunities offered
by the exotic isotope accelerator REX in conjunction with 
the modern Germanium $\gamma$ spectrometer MINIBALL.
\end{abstract}

\vspace{\hssp}
\section{Introduction}
\vspace{\hssp}
\pichskip{2mm}
\parpic(5cm,6.3cm)(0cm,5.6cm)[r]{\includegraphics[width=5cm]{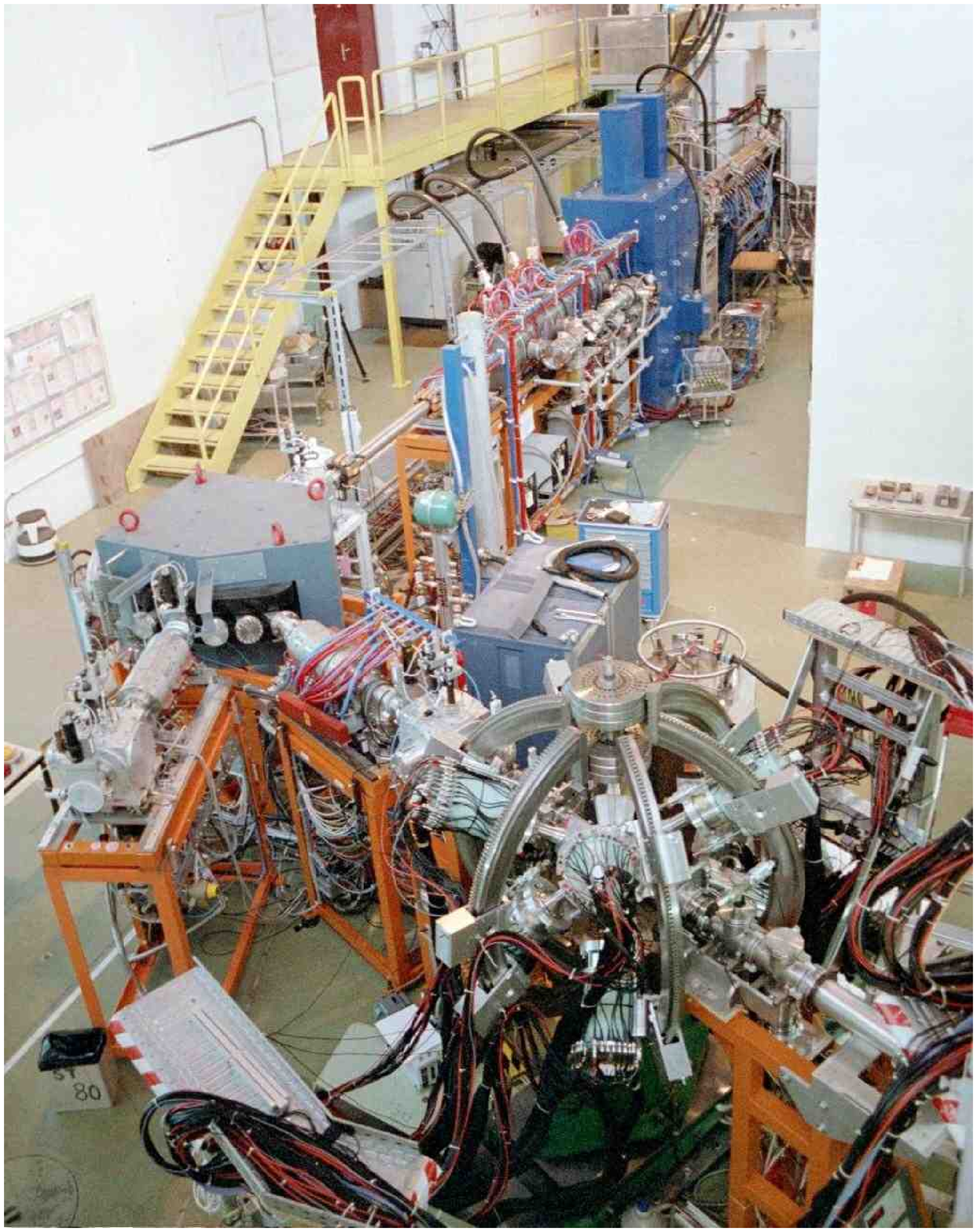}}
The Radioactive beam EXperiment (REX) \cite{hab01} at ISOLDE, CERN
is a pilot experiment to demonstrate a novel technique to 
bunch, charge breed and accelerate radioactive ions provided by
the ISOLDE facility.
The experiment was commissioned in 2002 and first physics experiments using the accelerated
exotic nuclei with a maximum beam energy of 2.2 MeV/u%
\footnote[1]{From spring 2004 onwards the maximum beam energy will be 3.1 MeV/u.}
were performed in 2003.
For further details on REX and its future see \cite{ced03}.
A picture of the setup in early 2002 is shown on the right
with the MINIBALL array on the 65$^\circ$ beamline in the foreground.
By the end of 2003 REX became a dedicated CERN user facility.

\section{Experimental Setup}

The main experimental device currently used with REX is the 
MINIBALL HP-Ge array \cite{ebe01,wei02}.
The array consists of 24 6-fold segmented, individually encapsulated, 
HP-Ge detectors arranged in 8 triple cryostats.
The detectors are mounted on an adjustable frame which allows for an easy adaptation of 
the geometry to the specific experimental requirements.
The central core and six segment electrodes of each detector
are equipped with a preamplifier with a cold stage and a warm main board.
The charge integrated signals are subsequently digitized (12 bit, 40 MHz)
and analyzed online and onboard the DGF-4C CAMAC card from XIA \cite{xia}.
Besides energy and timing information the (user) algorithms implemented on the card 
\cite{wei02,xia,gun00,lau99,lau04} determine the interaction point of each $\gamma$ ray 
in the detector via pulse shape analysis (PSA),
resulting in an about 100-fold increase in granularity in comparison 
to a non-segmented detector (the position resolution is about 7-8 mm FWHM, depending 
on the energy of the $\gamma$ ray).
All digitizers run independently (receiving the same clock signal)
and each single event is time stamped by
a 40 MHz clock for offline event reconstruction.

In addition to the MINIBALL array an annular charged particle detector telescope 
(of CD type, see \cite{dav03})
is employed consisting of a $\sim 500\;\mu$m thick $\Delta E$ followed by a 
$\sim 500\;\mu$m thick $E$ detector.
The $\Delta E$ detector is highly segmented (24x4 annular and 16x4 radial strips) to allow 
for a kinematic reconstruction of the events and covers an angular
range from $15^\circ$ to $50^\circ$.
A parallel plate avalanche counter \cite{cub00} was used to monitor the beam
at zero degrees without stopping it,
so that the (radioactive) beam particles are not deposited in the target area 
and reach the beam dump to reduce the radioactive decay background.

\begin{figure}
\centerline{\includegraphics[width=8cm,angle=-90]{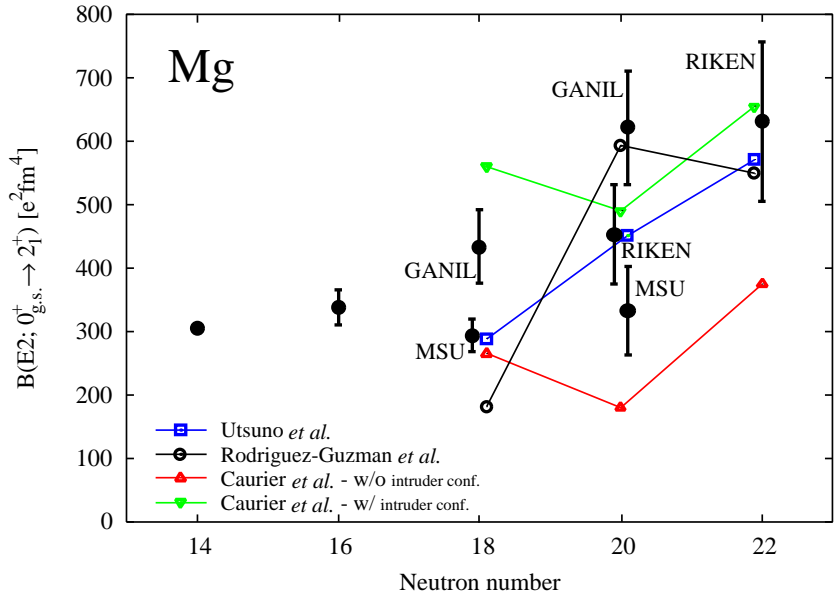}}
\caption{\label{fig:be2} $B(E2; 0_\mathrm{gs}^+ \rightarrow 2_1^+)$ values for the neutron-rich
even-even magnesium isotopes.
The references are:
Utsuno-\cite{uts99},
Rodriguez-Guzman-\cite{rod00},
Caurier-\cite{cau01},
GANIL-\cite{chi01},
MSU-\cite{pri99},
RIKEN-\cite{mot95,iwa01}.
See text for details.
}
\end{figure}

\section{First Experiments}
The primary aim of REX and MINIBALL is the investigation of the development of 
the structure of nuclei far from stability.
The reactions of choice to study the single particle and collective properties 
of nuclei with low-energy re-accelerated ISOL beams ($E_\mathrm{beam}\sim$ Coulomb barrier)
are single-nucleon transfer reactions and Coulomb excitation, respectively.
Especially favorable for neutron-rich nuclei are (d,p) reactions (in inverse kinematics) on
deuterated polyethylene targets, since here the cross sections are relatively high
\cite{len98} and the neutron-pickup product is more neutron-rich by one neutron.
By determining the differential cross sections to certain states as well as 
the angular distribution of the deexcitation $\gamma$ rays their spins
and parities can be determined.
From the absolute cross section, spectroscopic factors can be extracted.

An interesting region in the chart of nuclides is around the $N=20$ nucleus $^{32}$Mg. 
This region is well known as the island of deformation for almost 30 years \cite{kla69,thi75}, 
yet still, the knowledge on the nuclei in this region is very limited.  
Even today the extent of intruder ground state configurations 
in this region of the nuclear chart is not known.
$B(E2)$ values for the even-even nuclei up to $N=22$ ($^{34}$Mg) have been
measured \cite{mot95,pri99,chi01,iwa01} rather recently,
mainly by intermediate-energy Coulomb excitation
at projectile fragmentation facilities.
However, the data are very imprecise and measurements of different laboratories
disagree by as much as a factor of two
indicating that the systematic errors are not completely understood.
The main contribution to the systematic error in these experiments
is probably due to Coulomb--nuclear interference effects,
which are always present at such high beam energies, even if the scattering angle
of the projectiles is restricted to small values. 
In contrast, the Coulomb excitation studies planned at REX-ISOLDE will
be performed with beam energies well below the Coulomb barrier, yielding
truly model independent results.
In Fig.\ \ref{fig:be2} the theoretical and experimentally determined 
$B(E2; 0_{\mathrm{gs}} \rightarrow 2_1^+)$ for the
neutron-rich Mg isotopes as a function of neutron number $N$ are shown.
It should be noted that all the theoretical results 
\cite{uts99,rod00,cau01,ots02} are rather recent 
but give (nevertheless) very different values.
Based on the data shown, a discrimination between the theoretical models is
not possible.

Most information on the odd and odd-A nuclei stems from $\beta$ decay experiments
(see e.g. \cite{num01} and references therein)
and in the future it will be attempted to measure ground state nuclear moments by
the $\beta$ NMR technique after polarization with a LASER \cite{kei00}.

\begin{figure}
\centerline{\includegraphics[width=9cm]{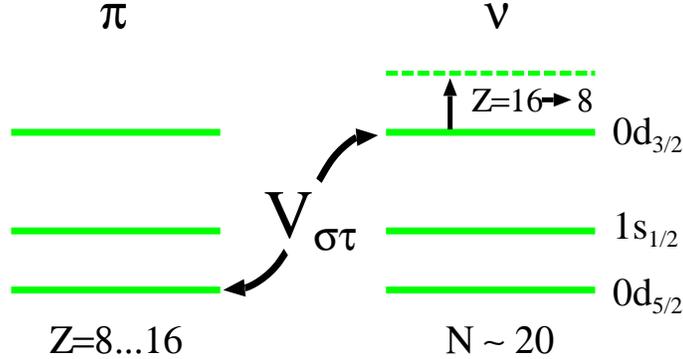}}
\caption{\label{fig:ots_sigma_tau} Illustration of the $V_{\sigma\tau}$ interaction between
neutron and proton orbitals with the same orbital angular momentum $l$
and different total angular momentum $j$.
This interaction was found to be very attractive for
stable nuclei and rather weak for exotic neutron-rich nuclei, 
where the corresponding proton orbitals are often not filled,
contributing to the appearance of new magic numbers \cite{ots02}.
}
\end{figure}
In a recent publication Otsuka \textit{et al.} \cite{ots02} pointed out the importance of the
$V_{\sigma\tau}$ (residual) interaction between orbits of the same orbital
angular momentum, but with different total angular momentum between neutron and proton
orbitals (see illustration in Fig. \ref{fig:ots_sigma_tau} for the sd-shell).
It is due to this interaction that the effective single particle energy (ESPE) 
of the $\nu$d$_\frac 32$ orbital is much lower in 
energy for stable nuclei, where the $\pi$d$_\frac 52$ orbital is nearly filled (resulting in a strong
attraction of the orbitals) 
in comparison to very neutron-rich nuclei, where the $\pi$d$_\frac 52$ is nearly empty.
$^{24}$O ($\pi$d$_\frac 52$ completely empty), e.g., 
shows features of a double magic nucleus
due to the rise of the ESPE of the $\nu$d$_\frac 32$ orbital\cite{ots02}.
To further test the role of $V_{\sigma\tau}$ in the structure of these neutron-rich nuclei
more precise experimental data are urgently needed.

At REX-ISOLDE with MINIBALL there is the possibility to study nuclei far from stability with standard
nuclear physics tools (in inverse kinematics), which are not only well proven, but which 
also allow a direct comparison of the experimental results to 
ones obtained with stable beams.
Therefore a program was started to systematically
study the neutron-rich nuclei in this region via neutron-pickup reactions
and Coulomb excitation.

\section{Preliminary Results}

\begin{figure}
\centerline{\includegraphics[width=8cm,angle=-90]{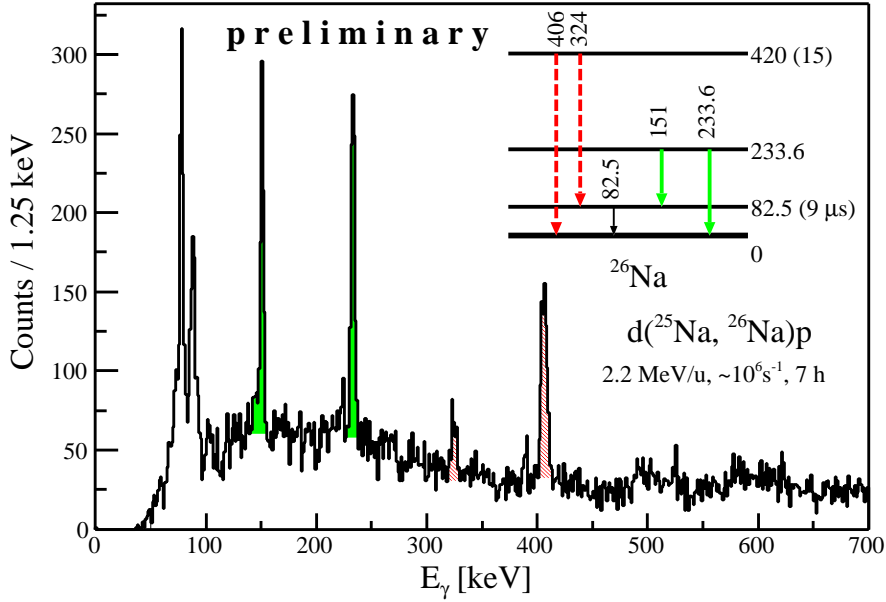}}
\caption{\label{fig:na26_spec}
$\gamma$ spectrum in coincidence with any signal in the $\Delta E$-$E$ detector.
Transitions in $^{26}$Na (3 neutrons away from stability) are marked.
See text for details.  The two lines around 80 keV are due to other reaction channels.
}
\end{figure}

\begin{figure}
\centerline{\includegraphics[width=8cm,angle=-90]{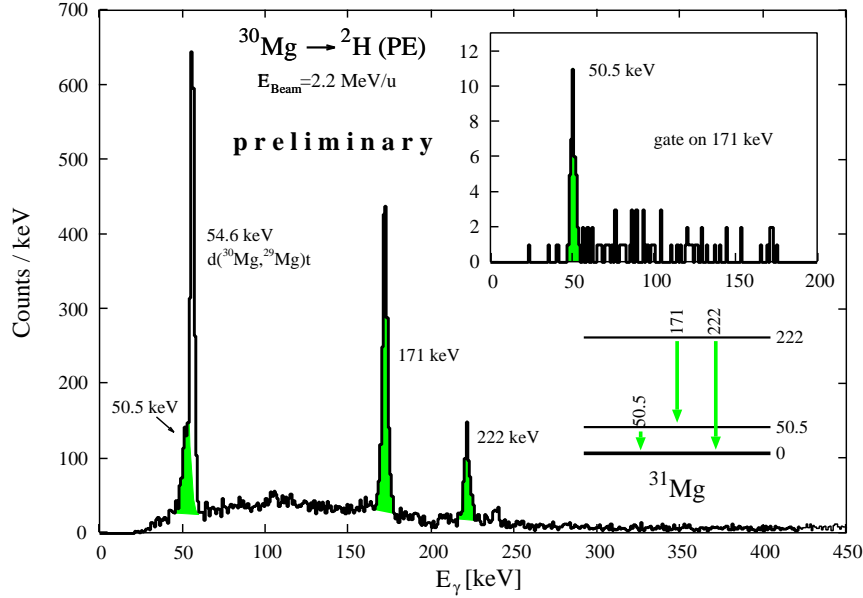}}
\caption{\label{fig:mg31_spec}
$\gamma$ spectrum coincident with any signal in the particle detector.
Please note the almost negligible background and the extremely low-energy threshold of only 40-50~keV.
The inset shows a $\gamma$ energy spectrum in coincidence with the 171~keV line.
As expected only the 50.5 keV transition is seen with still good statistics,
demonstrating the high efficiency of the MINIBALL array.
}
\end{figure}

Preliminary $\gamma$ energy spectra are shown in Figs.\ \ref{fig:na26_spec} and \ref{fig:mg31_spec}
to demonstrate the quality of the spectra measured with a radioactive beam.

The spectrum in Fig.\ \ref{fig:na26_spec} was taken in about 7 hours
with a $^{25}$Na beam (2 neutrons away from stability) 
with a beam energy of 2.2 MeV/u and an intensity of
about $10^6\;\mathrm s ^{-1}$ on a deuterated polyethylene target with a 
thickness of $10\;\mu\mathrm m$.
The transitions with energies of 151 and 234 keV depopulating the
state at 234 keV were already known.
In this low-energy part of the $\gamma$ spectrum two more lines can be identified and
assigned to $^{26}$Na: 324 and 406 keV.
The energy difference of the two transitions (82 keV) corresponds exactly
to the energy difference of the two known transitions, suggesting that
they depopulate the same state to the ground state (406~keV line) and to
the first excited state at 82 keV (324~keV line).
Furthermore a level with an energy of $(420\pm15)$~keV was previously known 
and the two transitions can therefore be assigned as shown in the inset in the
figure.

The two lines around 80~keV are due to other reaction channels.
The line at 78.1~keV results from a fusion-evaporation reaction of 
the projectile with the
$^{12}$C nuclei also present in the polyethylene target, leading to the population 
of the first excited state in $^{32}$P via the $^{12}$C($^{25}$Na,$\alpha$n)$^{32}$P
reaction.
Indeed this transition leads to the only line visible, when gating
on energies above 10~MeV in the CD $\Delta E$ detector, since
only particles with a $Z$ larger than one can deposit more energy.
The transitions with energies of 89.5 keV are due
to inelastic scattering of the $^{25}$Na beam nuclei
on the deuterons in the target (d($^{25}$Na,$^{25}$Na$^*$)d).

Fig.\ \ref{fig:mg31_spec} shows a spectrum taken in about 66 hours with a 
$^{30}$Mg beam (4 neutrons away from stability with a halflife of only 335 ms) 
with an intensity of about $2\cdot10^4\;\mathrm s^{-1}$ 
($E_{\mathrm{beam}}=2.2\;\mathrm{MeV/u}$) on the same target ($10\;\mu$m deuterated polyethylene).
Several known transitions in $^{31}$Mg can be seen, namely at 50.5 keV, 171 keV, and 222 keV.
In addition a strong line at 54.6 keV is evident.
This transition is coincident with tritons in the particle detector and corresponds
therefore to the transition of the first excited state in $^{29}$Mg (populated
via the d($^{30}$Mg,$^{29}$Mg)t reaction) to the ground state.
The inset shows a spectrum gated on the 171 keV line. 
As expected only the 50.5 keV transitions can be seen in this spectrum.
The statistics is still rather good demonstrating the high efficiency of
the MINIBALL array.
Please note the very low background seen in these spectra, even though the
beam particles are radioactive, and the extremely low-energy threshold of only 
40-50~keV, which should be compared to a threshold of typically several 
100 keV in experiments with fast beams.

\begin{figure}
\centerline{\includegraphics[width=6cm,angle=-90]{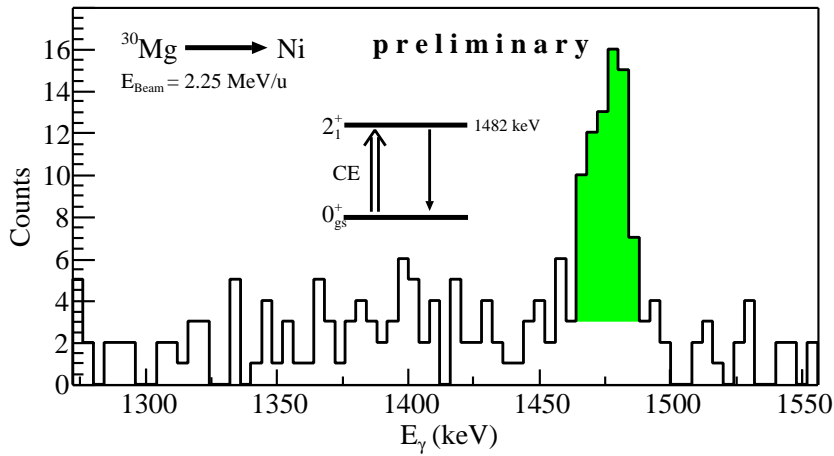}}
\caption{\label{fig:mg30_spec}
Coulomb excitation of $^{30}$Mg on a $^\mathrm{nat}$Ni target.  
The position information on the $\gamma$ rays and on the 
$^{30}$Mg nuclei registered in the CD detector was used for a preliminary Doppler correction.
Only events where the beam particles ($^{30}$Mg) were detected were included 
in the analysis (see text).}
\end{figure}
A preliminary spectrum from the Coulomb excitation of $^{30}$Mg on
a Ni target with a beam energy of 2.25~MeV/u is shown in Fig.\ \ref{fig:mg30_spec}.
The energies of the detected $\gamma$ rays were Doppler corrected using the position information
on the $\gamma$ rays from the MINIBALL array and the position
information on the $^{30}$Mg particles detected in the CD detector;
only preliminary values for the geometry were available
when this spectrum was generated resulting in a resolution of only about 20 keV FWHM.
Only events with $^{30}$Mg nuclei detected in the CD detector were
included in the analysis; events where the recoiling Ni nuclei were
registered still have to be analyzed, increasing the
shown statistics by a factor of about two.
Simultaneously to the excitation of the beam particles also the target nuclei are
Coulomb excited.  
When the proper Doppler correction for the recoiling target nuclei
is performed 
transitions due to Coulomb excitations of $^{58,60}$Ni with energies of 
1454~keV and 1332~keV, respectively, show up as sharp peaks in the spectrum.
The lifetimes of the first excited $2^+$ states of $^{58,60}$Ni are very well known and 
can therefore serve as a calibration for the extraction of the 
$B(E2;  0_\mathrm{gs}^+ \rightarrow 2_1^+)$ of $^{30}$Mg.
After the final analysis the statistical uncertainty on the extracted $B(E2)$
should be on the order of 7-9\% which is comparable to the total error of the
MSU result \cite{pri99}.
The systematic uncertainties will be much smaller, though.

In conclusion, REX and MINIBALL are fully operational and first physics experiments
focusing on the nuclear structure of the neutron-rich Na and Mg isotopes were performed.
A preliminary analysis shows the high quality of the data that can be obtained with MINIBALL
at REX-ISOLDE with a radioactive, low-energy and low-intensity beam.
The data are currently under analysis.
  
This work was supported by the German Bundesministerium f\"ur Bildung und Forschung.


\end{document}